\documentclass[lettersize,journal]{IEEEtran}
\usepackage{amsmath,amsfonts}
\usepackage{algorithmic}
\usepackage{algorithm}
\usepackage{array}
\usepackage[caption=false,font=footnotesize,labelfont=sf,textfont=sf]{subfig}
\usepackage{textcomp}
\usepackage{stfloats}
\usepackage{url}
\usepackage{verbatim}
\usepackage{graphicx}
\usepackage{cite}
\usepackage{siunitx}
\usepackage{booktabs}

\begin{document}
\bstctlcite{IEEEexample:BSTcontrol}

\title{Modeling and Testing Superconducting Artificial CPW Lines Suitable for Parametric Amplification}

\author{F.~P.~Mena, D.~Valenzuela, C.~Espinoza, F.~Pizarro, B.-K.~Tan, D.~J.~Thoen, J.~J.~A.~Baselmans, R.~Finger
\thanks{F.~P.~Mena and C.~Espinoza are with the Central Development Laboratory, National Radio Astronomy Observatory, 1180 Boxwood Estate Rd, Charlottesville, VA 22903.}%
\thanks{D.~Valenzuela is with the Electrical Engineering Department, Faculty of Physical and Mathematical Sciences, University of Chile, Av. Tupper 2007, Santiago, Chile.}%
\thanks{F.~Pizarro is with the Escuela de Ingeniería Electrica, Pontificia Universidad Católica de Valparaíso, Valparaiso 2362804, Chile.}%
\thanks{B.-K.~Tan is with the Department of Physics (Astrophysics), University of Oxford, Denys Wilkinson Building, Keble Road, Oxford, OX1 3RH, UK}%
\thanks{D.~H.~Thoen and J.~J.~A.~Baselmans are with the Netherlands Institute for Space Research (SRON), Instrument Science Group - Litho, Niels Bohrweg 4, 2333 CA Leiden, The Netherlands.}%
\thanks{J.~J.~A.~Baselmans is also with the Department of Electrical Engineering, Faculty of Mathematics and Computer Science (EEMCS), Delft University of Technology, Mekelweg 4, 2628 CD Delft, The Netherlands.}%
\thanks{R.~Finger is with the Astronomy Department, Faculty of Physical and Mathematical Sciences, University of Chile, Camino El Observatorio 1515, Santiago, Chile.}%
}

\markboth{Journal of \LaTeX\ Class Files,~Vol.~14, No.~8, August~2021}%
{Shell \MakeLowercase{\textit{et al.}}: A Sample Article Using IEEEtran.cls for IEEE Journals}

\IEEEpubid{0000--0000/00\$00.00~\copyright~2021 IEEE}

\maketitle

\begin{abstract}
Achieving amplification with high gain and quantum-limited noise is a difficult problem to solve. 
Parametric amplification using a superconducting transmission line with high kinetic inductance is a promising technology not only to solve this problem but also adding several benefits. When compared with other technologies, they have the potential of improving power saturation, achieving larger fractional bandwidths and operating at higher frequencies. In this type of amplifiers, selecting the proper transmission line is a key element in their design. Given current fabrication limitations, traditional lines such as coplanar waveguides (CPW), are not ideal for this purpose since it is difficult to make them with the proper characteristic impedance for good matching and slow-enough phase velocity for making them more compact. Capacitively-loaded lines, also known as artificial lines, are a good solution to this problem. However, few design rules or models have been presented to guide their accurate design. This fact is even more crucial considering that they are usually fabricated in the form of Floquet lines that have to be designed carefully to suppress undesired harmonics appearing in the parametric process. In this article we present, firstly, a new modelling strategy, based on the use of electromagnetic-simulation software, and, secondly, a first-principles model that facilitate and speed the design of CPW  artificial lines and of Floquet lines made out of them. Then, we present comparisons with experimental results that demonstrate their accuracy. Finally, the theoretical model allows to predict the high-frequency behaviour of the artificial lines showing that they are good candidates for implementing parametric amplifiers above 100~GHz.
\end{abstract}


\section{Introduction}\label{sec:intro}
\IEEEPARstart{A}{mplification} of small signals and limiting the excess noise introduced by this process are ubiquitous needs of modern electronics. In some state-of-the-art applications, such as quantum computing or astronomical instrumentation, the noise introduced by the amplifier has to be close to the limit imposed by quantum mechanics~\cite{Bardin2021}. Achieving amplification with high gain and quantum-limited noise is a difficult problem to solve, which is even exacerbated if operation at high frequencies or large fractional bandwidths, or both, are required. Semiconductor technology, in the form of HEMT-based amplifiers, has played an important role in filling this need~\cite{Bardin2021,Pospieszalski2018}. However, the minimum noise and fractional bandwidth they can achieve is still limited, particularly at operational frequencies above around \SI{100}{\giga\hertz}~\cite{Pospieszalski2017}. One approach to solve this problem is the use of superconducting travelling-wave parametric amplifiers that can be implements in two ways, as an array of Josephson junctions~\cite{Aumentado2020} or SQUIDs~\cite{Zorin2016}, or as a long highly non-linear transmission line~\cite{Eom2012}. The former implentation, on the one hand, has demonstrated quantum-limited noise at low frequencies (below \SI{10}{\giga\hertz}) but can handle very limited powers making it unfit for some applications~\cite{Aumentado2020,Esposito2021}. On the other hand, amplifiers based on long superconducting transmission lines, known as traveling-wave kinetic-inductance parametric amplifiers (TKIPAs), have also demonstrated quantum-limited noise~\cite{Malnou2021,klimovich2023} but have the potential advantage of being able to handle larger input powers and operating at much larger frequencies, limited only by the energy gap of the superconductor employed to fabricate them~\cite{Eom2012}.

The implementation of 
TKIPAs requires selecting the appropriate transmission-line geometry. The first demonstration of such amplifiers used bare CPW lines since they are relatively easy to implement \cite{Eom2012}. Unfortunately, given the large inductance of the superconductor, due to its kinetic inductance, and limitations in the geometrical dimensions that can be patterned, only lines with rather large characteristic impedances (above \SI{150}{\ohm}) can be fabricated, making its matching with external circuits difficult. Furthermore, the achieved phase velocity (around $0.1 c$, where $c$ is the speed of light) implies that a long line is needed in order to obtain a reasonable gain. Capacitively-loaded superconducting microstrip and CPW lines have been investigated to overcome these problems~\cite{Adamyan2016,Chaudhuri2017,klimovich2023}. Artificial lines implemented using such geometries have the potential to achieve lower impedances and phase velocities below $0.05c$ under reasonable fabrication constraints.
\IEEEpubidadjcol

The easiest way to implement the capacitive loads is in the form of straight stubs, similar to interdigital capacitances~\cite{Chaudhuri2017,klimovich2023,Yoon2003}. If the line is, moreover, fabricated in a CPW geometry, the result is a versatile line which is relatively easy to fabricate. Although the design of this type of transmission lines is not trivial, few attempts have been made to provide first-principle models or methods to properly guide their design~\cite{Nakagawa2020,Malnou2021,Tan2022}. 
Filling the need of an adequate and fast modelling tool is even more important taking into account that, to use them in parametric amplifiers, they are usually implemented in the form of periodically-repeating unit cells (Floquet lines) that need to be carefully designed to achieve proper phase matching and suppress the main unwanted harmonics appearing in the parametric process~\cite{Eom2012}. Moreover, an accurate design would allow to implement lines with a precise characteristic impedance that could minimize the large gain ripples seen in travelling-wave parametric amplifiers.

In this work we present two new design approaches and demonstrate that they can describe in an excellent fashion the transmission measurements of a set of Floquet lines made out of segments of superconducting CPW artificial lines. One approach is based in the use of electromagnetic simulation software that, at variance with other software-aided methods, does not need large computational resources, converting it in a versatile tool for aiding in design. The second approach is a first-principles model that combines a circuital model with transfer-matrix theory. Besides being a helpful design tool, the model also demonstrates the possible use of CPW artificial lines at frequencies above \SI{100}{\giga\hertz}.

\section{Geometry of the CPW Artificial Line}\label{sec:artificial}

\begin{figure}[t]%
    \centering
    \subfloat[\centering ]{{\includegraphics[width=0.52\linewidth]{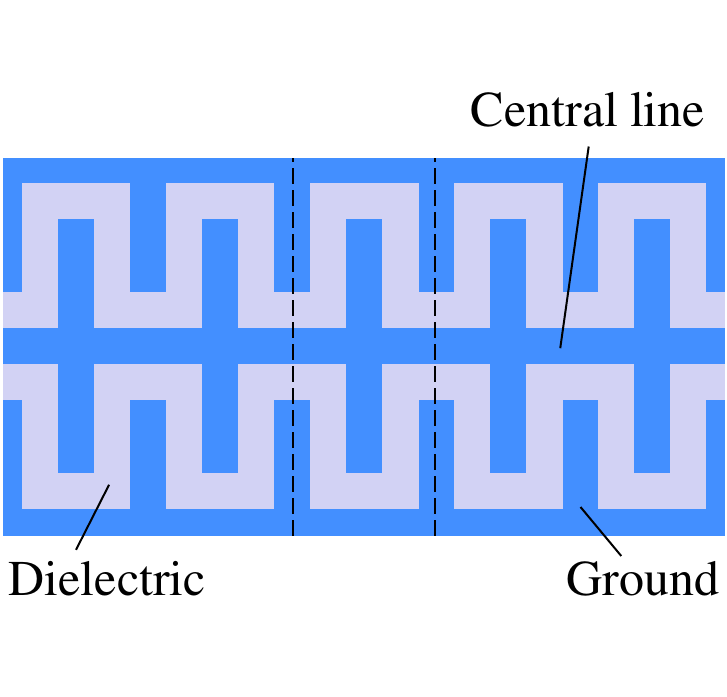} }}%
    \quad
    \subfloat[\centering ]{{\includegraphics[width=0.36\linewidth]{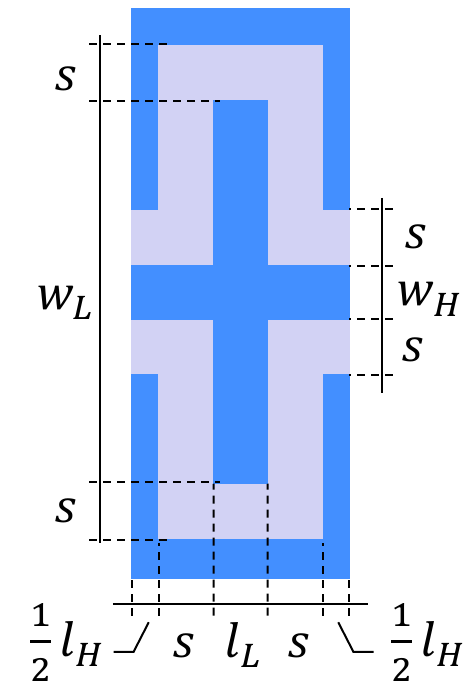} }}%
    \caption{(a)~Capacitively-loaded artificial CPW transmission line. Vertical dashed lines indicate one unit cell. (b)~Zoom of the unit cell showing the dimensions that define it.}%
    \label{fig:artificial}%
\end{figure}


The layout of the superconducting CPW artificial line studied in this work is presented in Fig.~\ref{fig:artificial}. It consists of a CPW line where straight fingers, that form capacitances with the ground, are appended perpendicularly to it. In this way, additional capacitance per unit length is added to the CPW line in order to compensate for its large kinetic inductance. In principle, thus, the length of the fingers can be used to tune the impedance and phase velocity of the line. It is for this reason that it is very important to have an accurate methodology that allows selecting their correct length for a given purpose.


\section{Floquet Lines with CPW Artificial Lines}\label{sec:Floquet}

\subsection{Concept and Basic Equations}\label{sec:Floquet-concept}

\begin{figure}[!t]
    \centering
    \includegraphics[width=\linewidth]{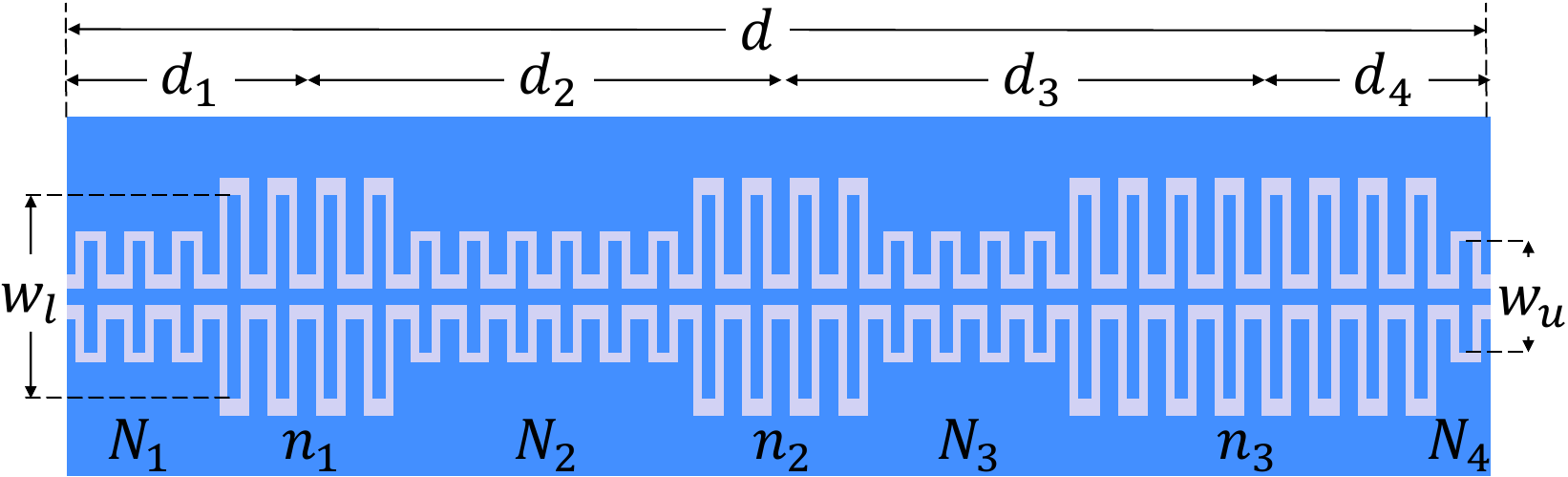}
    \caption{Unit cell of a Floquet transmission line made out of artificial CPW lines. $N_i$ and $n_j$ represent the number of fingers in each segment.}
    \label{fig:Floquet}
\end{figure}

A Floquet line (for a review see~\cite{Perruisseau2007}) consists of unit cells that repeat periodically. In that sense, the artificial line presented in the previous section is also a Floquet line. However, to avoid confusions, here we give this name to lines made out of segments of CPW artificial lines. A scheme of the unit cell of such a Floquet line is presented in Fig.~\ref{fig:Floquet}.

The properties of a Floquet line are completely determined by its unit cell. Indeed, let $T_{UC}$ be the transmission matrix of the latter,
\begin{equation*}
    T_{UC} =
    \begin{pmatrix} 
        A & B \\
        C & D 
    \end{pmatrix}.
\end{equation*}
Then, if $d$ is the length of the unit cell, the propagation constant, $\gamma$, and Bloch impedance, $Z_B$, of the entire Floquet line can be calculated from
\begin{equation}\label{eq:gamma}
    \cosh{\gamma d} = \frac{A+D}{2},
\end{equation}
and
\begin{equation}\label{eq:ZB}
    Z_B = \pm \frac{B}{\sqrt{A^2-1}},
\end{equation}
respectively. Moreover, the transmission of a Floquet line with $n$ unit cells is given by

\begin{equation}\label{eq:S21}
    S_{21} = \frac{2 Z_0 Z_B}{\left(Z_0^2+Z_B^2\right) \sinh{n \gamma  d}+2 Z_0 Z_B \cosh{n \gamma  d}},
\end{equation}
where $Z_0$ is the impedance to which it is connected.

\subsection{Fabrication}\label{sec:Floquet-fab}
\begin{figure}[!t]
    \centering
    \includegraphics[width=0.7\linewidth]{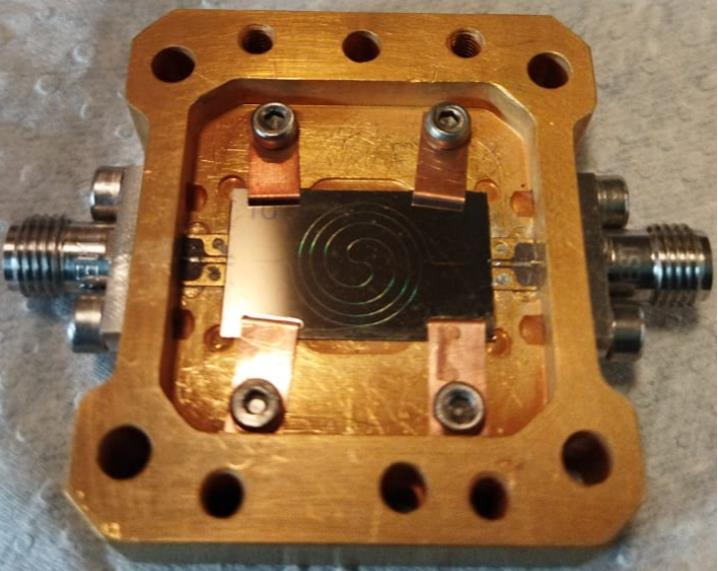}
    \caption{Superconducting Floquet line mounted for characterization. The dimensions of the chip are $\SI{20}{\milli\meter} \times \SI{12}{\milli\meter}$.}
    \label{fig:mounted}
\end{figure}

\begin{table*}[t]
    \centering
    \caption{Parameters defining the characterized Floquet lines.}
    \label{tab:dimensions}
    \vspace{-5mm}
    \begin{tabular}[t]{cccccccccccc}
        \toprule
        ID & $s$ & $w_u$  & $w_l$   & $n_1$  & $n_2$  & $n_3$  & $N_1$  & $N_2$  & $N_3$  & $N_4$   & $n$  \\
        -- & \si{\micro\meter} & \si{\micro\meter}  & \si{\micro\meter}   & --  & --  & --  & --  & --  & --  & --  & --  \\
        \midrule
        A & 1   & 19   & 35   & 40   & 40   & 80  & 110   & 221   & 201   & 93   & 62   \\
        B & 2   & 26   & 40   & 20   & 20   & 40   & 38   & 77   & 67   & 29   & 40   \\
        \bottomrule
    \end{tabular}
\end{table*}%
Using an earlier version of the model to be presented presented in Sec.~\ref{sec:model} (see \cite{Valenzuela2020}), we designed and fabricated a set of the Floquet lines described above. We used Si ($\epsilon_r = 11.44$) as substrate and a 60-\si{\nano\meter} superconducting layer of
NbTiN deposited by the sputtering system LLS801 in static mode ($T_c = 14.7$~\si{\kelvin}, $\rho_n = 132$~\si{\micro\ohm\centi\meter}) \cite{Thoen2017}. Several versions were fabricated, all of them keeping $s=l_H=l_L=w_H$. The dimensions of the fabricated versions that are described in this article are given in Table~\ref{tab:dimensions}. A picture of one of the resulting devices mounted in a block for characterization can be seen in Fig.~\ref{fig:mounted}.

\subsection{Characterization}\label{sec:Floquet-character}

\begin{figure}[t]%
    \centering
    \subfloat{{\includegraphics[width=0.78\linewidth, clip, trim={0 20mm 0 0}]{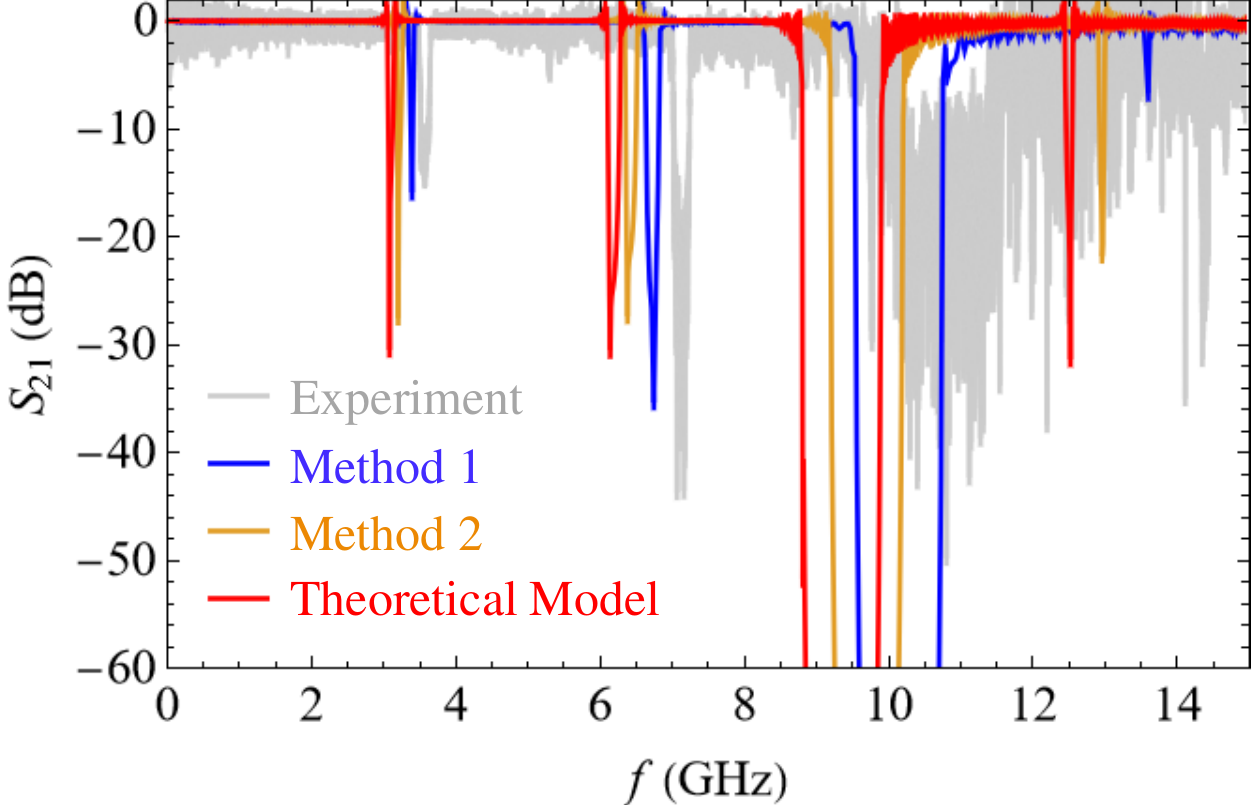} }}\\%
    \vspace{-3.5mm}
    \subfloat{{\includegraphics[width=.78\linewidth]{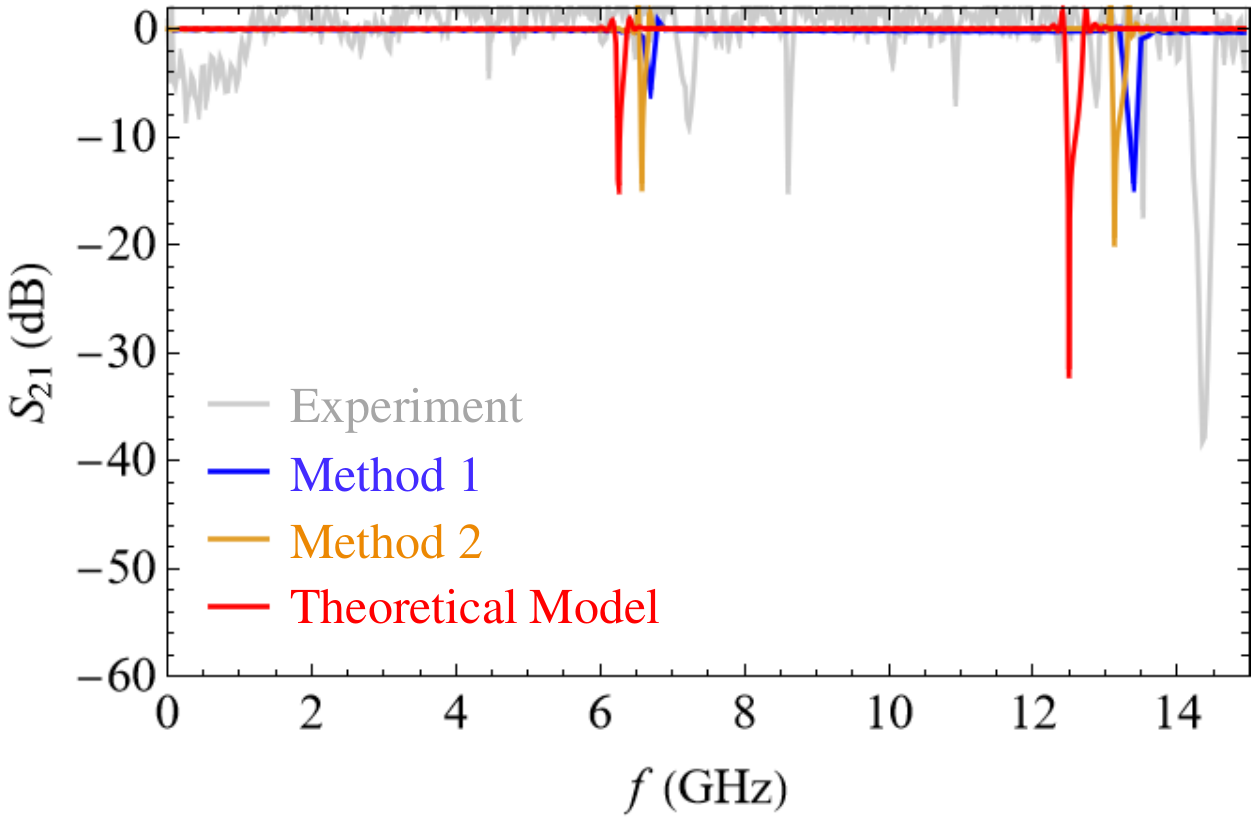} }}%
    \caption{Measured (gray line) and modelled transmissions of the Floquet lines presented in Table~\ref{tab:dimensions}. For modelling we used software aided methods~1 (blue line) and~2 (orange line), and the first-principles model (red line). (Top)~Device~A characterized at \SI{100}{\milli\kelvin}. (Bottom)~Device~B characterized at \SI{4}{\kelvin}. The measurements were made in different setups.}%
    \label{fig:measure}%
\end{figure}

Some of the fabricated devices were characterized at \SI{4}{\kelvin} and \SI{100}{\milli\kelvin} in two different setups. Two examples of those measurements are presented in the grey lines of Fig.~\ref{fig:measure}. To normalize the transmission a second set of measurements were made where the mounted device was replaced by a coaxial barrel connector.


\section{Modelling Floquet Lines with the Help of Electromagnetic Simulation Software}\label{sec:hfss}

\begin{figure}[t]%
    \centering
    \subfloat[]{{\includegraphics[width=0.95\linewidth]{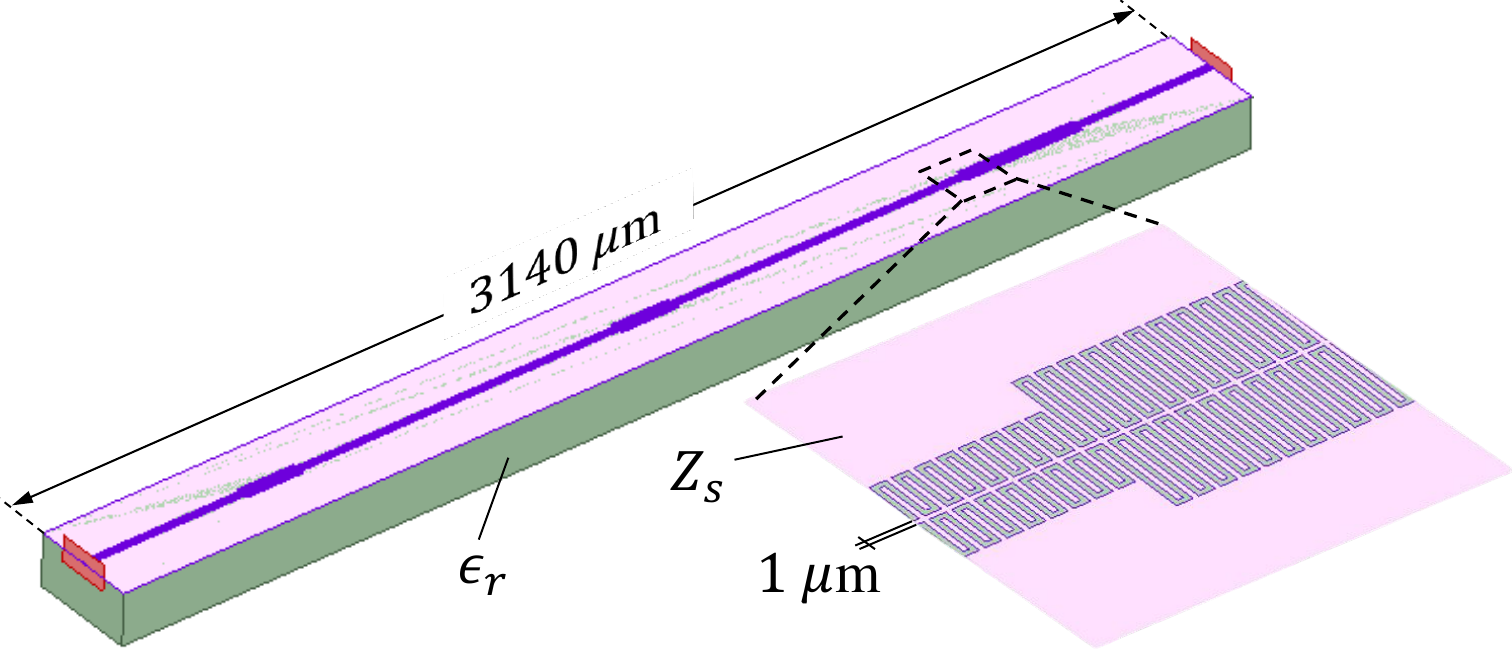} }}\\%
    \subfloat[]{{\includegraphics[width=.9\linewidth]{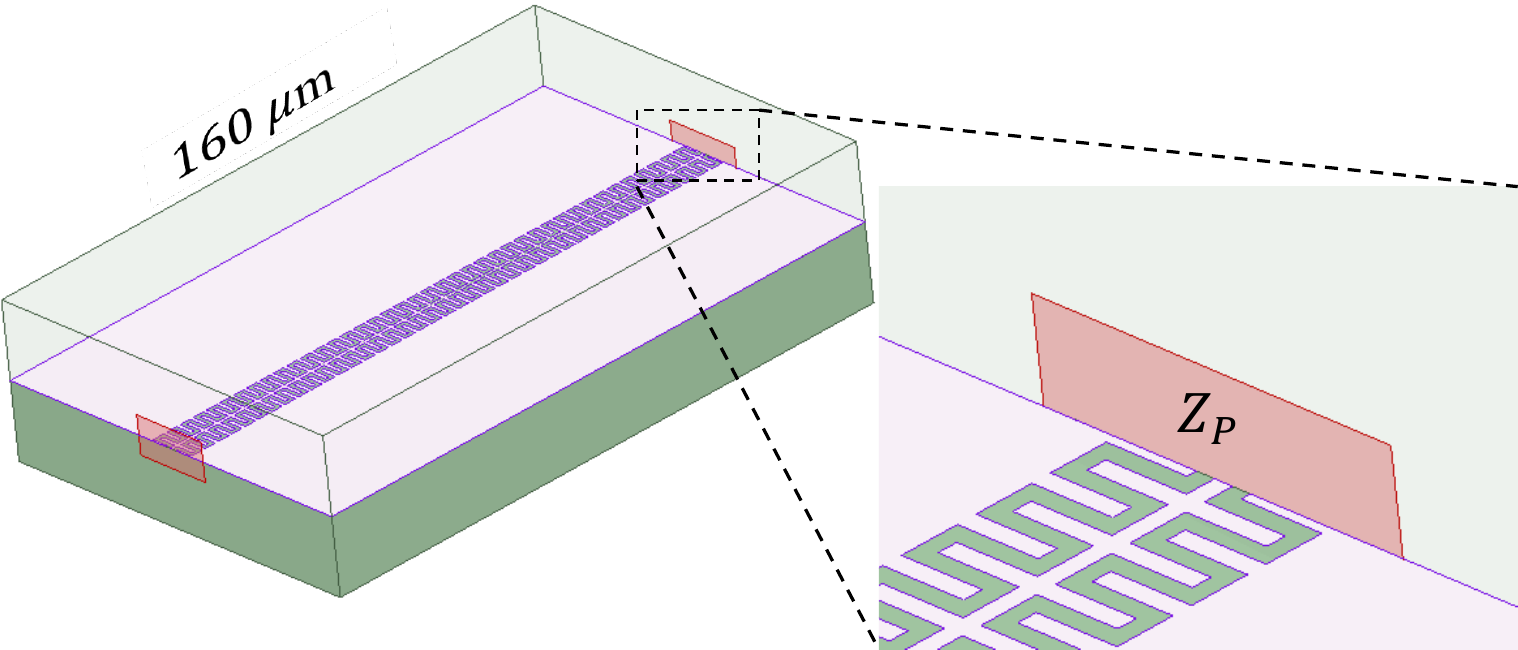} }}%
    \caption{Examples of CAD models used in the simulations with HFSS. The dielectric has a permittivity $\epsilon_r$ and the superconductor a surface impedance $Z_S$. The insets show the smallest features of the models. (a) The unit cell of device~A. (b) One of the simulated CPW artificial lines. The excitation port of the model has an impedance~$Z_P$.}%
    \label{fig:model}%
\end{figure}

In this section we present two different ways of modelling the Floquet lines. Both of them use the results of a commercially available electromagnetic simulation software, HFSS in this case, as an intermediate step. The CAD models used in both methods are presented in Fig.~\ref{fig:model}. The superconducting layer was modelled by assigning to it a surface impedance given by
\begin{equation*}
    Z_s=2\pi j f \mu_0 \lambda \coth{t/\lambda},
\end{equation*}
where $f$ is the frequency, $\mu_0$ is the vaccuum permeability, $t$ is the thickness of the superconductor, and $\lambda$ is its penetration depth~\cite{Kerr1999}. 

\subsection{Method 1: Using the Simulation of a Unit Cell}\label{sec:hfss1}
Following~\cite{Tan2022}, this method starts by simulating in HFSS one unit cell of the Floquet line. Then, the resulting S-matrix is converted to the transmission matrix from where its propagation constant, Bloch impedance, and transmission are calculated using (\ref{eq:gamma}), (\ref{eq:ZB}), and (\ref{eq:S21}), respectively. The transmission of devices~A and~B obtained after applying this method are presented in the blue lines of Fig.~\ref{fig:measure}. The agreement between the method and measurements is good, validating the former. Further simulations, not presented here, demonstrate that decreasing the convergence criterion improves the agreement.

Unfortunately, there is one major disadvantage of using this method. Given the large dimensions of the unit cell respect to its smallest features, it demands a large amount of computational resources which, in turn, makes it impractical for design. For example, the simulation of the unit cell of device~A, with a convergence criterion of $\Delta S = 0.1$, needed to obtain the results presented in Fig.~\ref{fig:measure}(a), took approximately \SI{30}{\minute} per frequency point using a desktop with a processor Intel Xeon W-2295 running at \SI{3.00}{\giga\hertz} and a RAM of 128~GB.

\subsection{Method 2: Using the Simulation of CPW Artificial Lines}\label{sec:hfss2}
\begin{figure}[t]%
    \centering
    \subfloat{{\includegraphics[width=0.78\linewidth, clip, trim={0 20mm 0 0}]{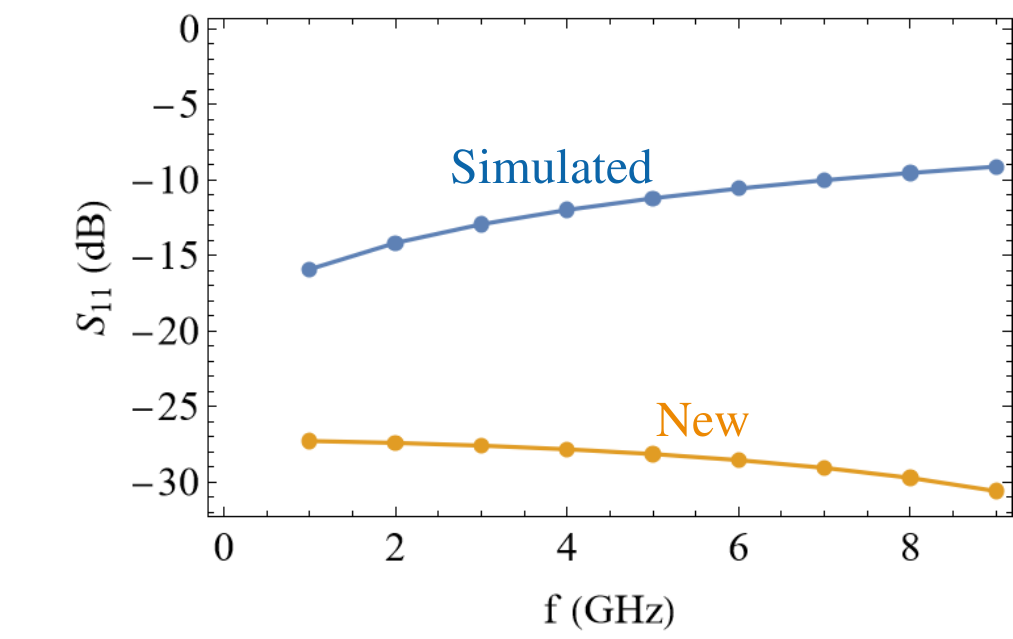} }}\\%
    \vspace{-3.7mm}
    \subfloat{{\includegraphics[width=.78\linewidth, clip, trim={0 19.5mm 0 0}]{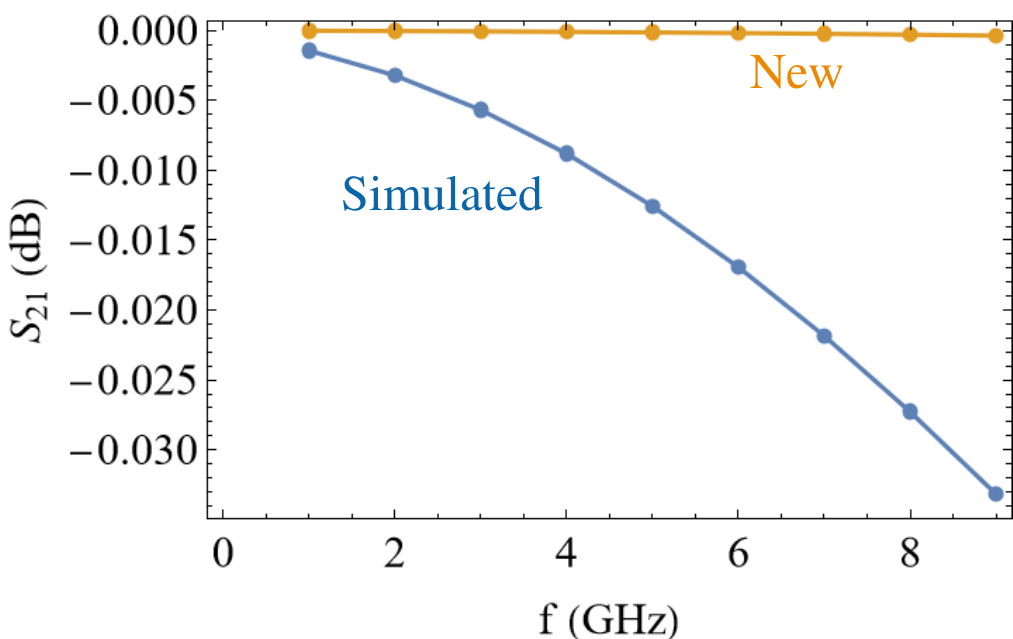} }}\\%
    \vspace{-3.7mm}
    \subfloat{{\includegraphics[width=.78\linewidth]{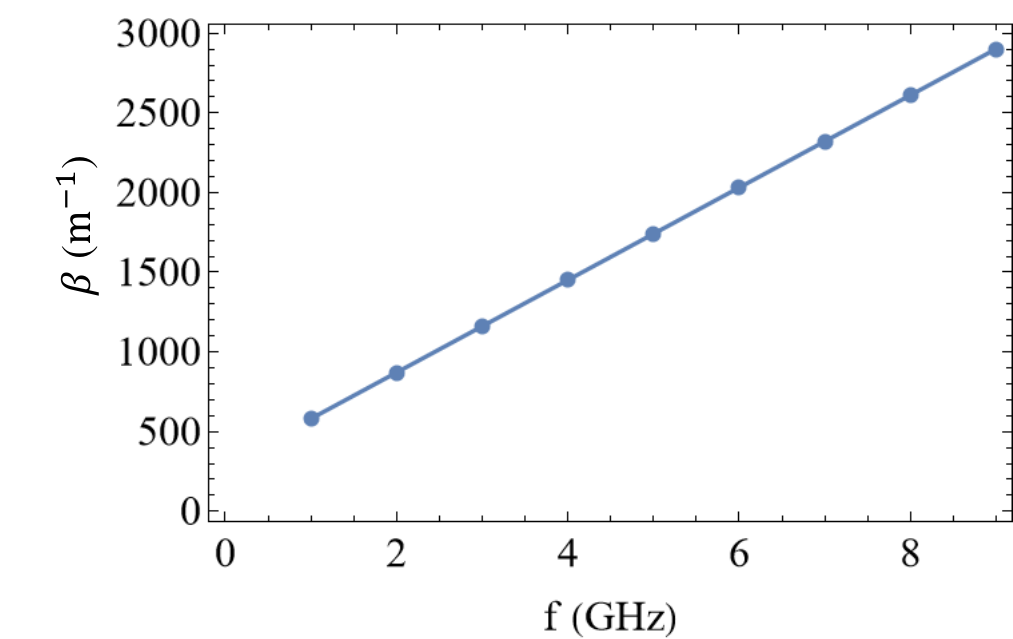} }}
    \caption{Example of applying software-aided method~2 to an artificial CPW line with $s=l_H=l_L=w_H=\SI{1}{\micro\meter}$ and $w_L=\SI{19}{\micro\meter}$, i.e. the central line of device~A. (Top)~Reflection before (blue line) and after (orange line) applying renormalization. (Middle)~Transmission before (blue line) and after (orange line) applying renormalization. (Bottom) Imaginary part of the propagation constant. The slope of the line correspond to a phase velocity of~$v=0.07c$.}%
    \label{fig:method2}%
\end{figure}%
If the transmission matrix,
\begin{equation}\label{eq:Ti}
    T_i =
    \begin{pmatrix} 
        \sinh{\gamma_i l_i} & Z_{0,i}\cosh{\gamma_i l_i} \\
        \frac{1}{Z_{0,i}}\cosh{\gamma_i l_i} & \sinh{\gamma_i l_i} 
    \end{pmatrix},
\end{equation}%
of every segment forming the unit cell of a Floquet line is known, they can be multiplied to obtain the transmission matrix of the unit cell. Having this matrix, the properties of the Floquet line are completely determined. 

Having demonstrated in the previous subsection the good accuracy of HFSS in describing Floquet superconducting lines, we have used it to determine the propagation constants and characteristic impedances needed to construct the matrices~(\ref{eq:Ti}). The method consist of two steps.

First, we need to realize that HFSS, or any other electromagnetic software used in the way described here, cannot give directly the impedance of any artificial line, $Z_{0}$, as the one depicted in Fig.~\ref{fig:model}(b). In fact, the impedance of the port, $Z_P$, calculated by HFSS corresponds to that of the small CPW section connected directly to it. If $S$ is the $S$-matrix given by HFSS, we assert that $Z_0$ corresponds to the impedance that minimizes $S_{\mbox{new},11}$ in the renormalization expression
\begin{equation*}
    S_{\mbox{new}} = A^{-1} \cdot (S-R) \cdot (I-R \cdot S)^{-1}\cdot A,
\end{equation*}%
where $I$ is the unit matrix, $A=\frac{\sqrt{Z_0 - Z_P}}{Z_0 + Z_P}I$, and $R=\frac{Z_0 - Z_P}{Z_0 + Z_P}I$. 

Second, after converting $S_{\mbox{new}}$ to a transmission matrix, it is possible to obtain the propagation constant according to (\ref{eq:Ti}). An example of this two-step process is presented in Fig.~\ref{fig:method2}. Importantly, although the minimization was with respect to the parameter $S_{11}$, it also maximizes $S_{21}$.

\begin{figure}[t]%
    \centering
    \subfloat{{\includegraphics[width=0.8\linewidth, clip, trim={0 17.5mm 0 0}]{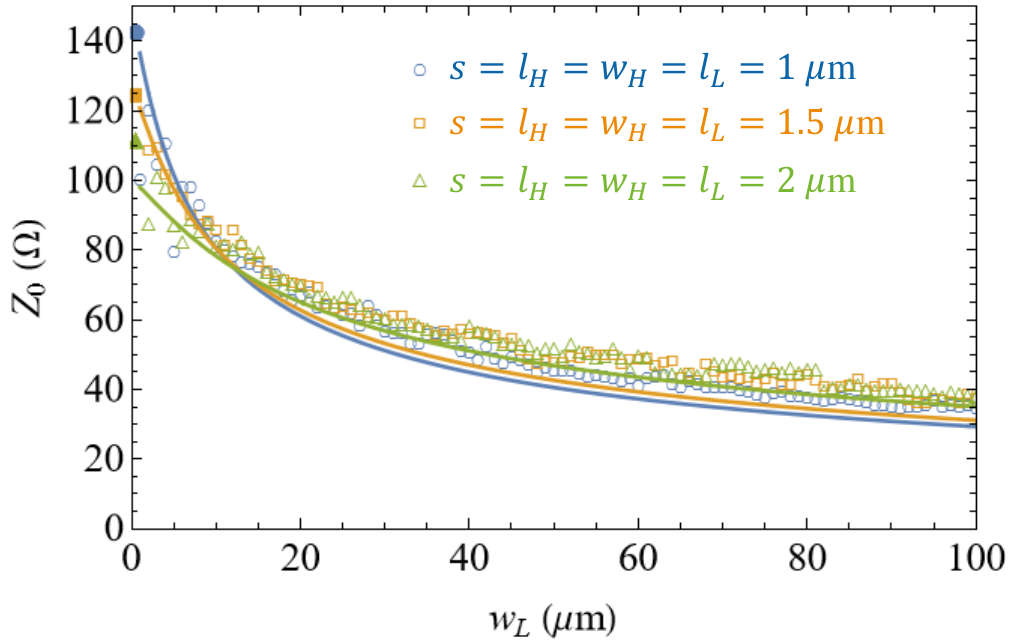} }}\\%
    \vspace{-3.5mm}    \subfloat{{\includegraphics[width=.8\linewidth]{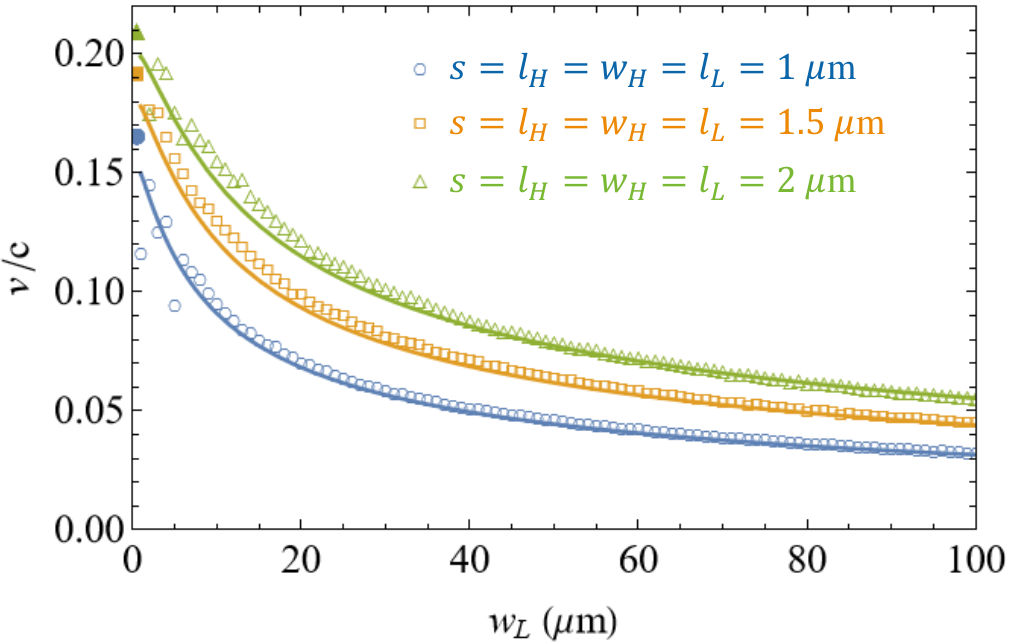} }}%
    \caption{Dependence of the properties of a CPW artificial line  with $w_L$ (see Fig.~\ref{fig:artificial}). This dependence was obtained using software-aided method~2 (empty symbols) and the first-principles model (lines). Solid symbols on the far left are the theoretical values of a CPW line, calculated according reference~\cite{Watanabe1994}, that has the dimensions of the central line, i.e. when $w_L \rightarrow 0$. (Top)~Characteristic impedance. (Bottom)~Phase velocity as fraction of the speed of light.}%
    \label{fig:method2b}%
\end{figure}%

For designing a CPW artificial line, one has to realize that, if the internal dimensions are kept constant, its properties can be tuned by varying $w_L$. Given the much less demanding computational resources needed for the required simulations, this method can be applied to obtain the necessary information, as presented in the empty symbols of Fig.~\ref{fig:method2b}. Remarkably, when $w_L \rightarrow w_H$, the values obtained by method~2 tend to those calculated for a CPW line with the dimensions of the central line. 
Figure~\ref{fig:method2b} also demonstrates that it is indeed possible to obtain 50-\si{\ohm} lines with this geometry together with a low phase velocity.

Using the values obtained in Fig.~\ref{fig:method2b}, the transmissions of the constructed devices were calculated and are presented in the orange lines of Fig.~\ref{fig:measure}. Agreement with measurements is also good but it can be improved if the convergence criterion used in the simulations of Fig.~\ref{fig:method2b} is diminished. The remaining small deviations between models~1 and~2 may be due to two factors. One possibility is that an electromagnetic simulation of the entire unit cell captures small interactions between loads and the central line that are not taken into account during multiplication of the transmission matrices. Another possibility is that, as Fig.~\ref{fig:method2b} shows, method~2 has more uncertainties when the values of $w_L$ are small. Indeed, additional simulations for lines using capacitive loads with $w_L>\SI{30}{\micro\meter}$, demonstrate that the difference between the two methods is smaller.


\section{First-principles Model of a CPW Artificial Line}\label{sec:model}
\subsection{Description of the Model}
\begin{figure}[t]%
    \centering
    \subfloat[]{{\includegraphics[width=0.4\linewidth]{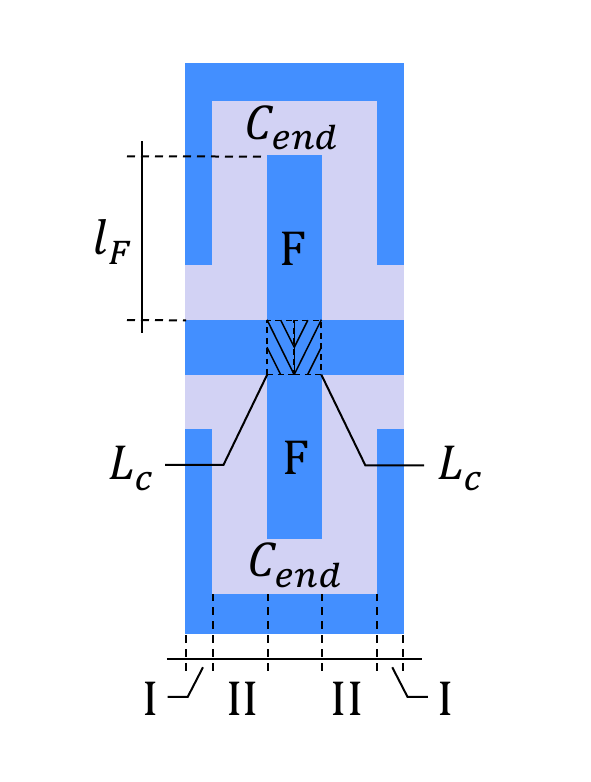} }}%
    \qquad
    \subfloat[]{{\includegraphics[width=0.4\linewidth]{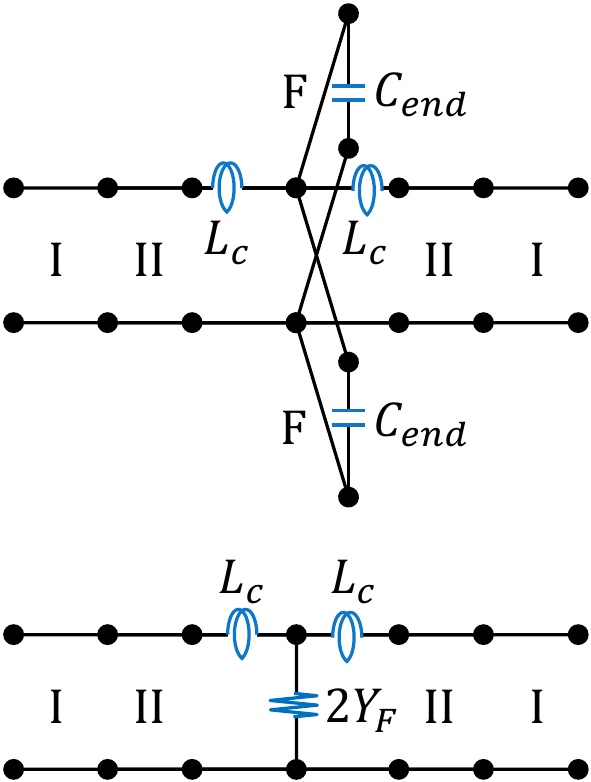} }}%
    \caption{(a)~Unit cell of the artificial CPW transmission line identifying the elements that are included in the equivalent circuit model: CPW lines I, II, and F, capacitance at the end of the capacitive load, $C_{end}$, and a free-standing superconductor of inductance $L_c$. Line F has a length $l_F = \frac{1}{2}(w_L-w_H)$. (b)~The top panel presents the corresponding circuit model which can be reduced to the one presented in the bottom panel.}%
    \label{fig:Tmodel}%
\end{figure}%
The model presented here shares some similarities with work presented before~\cite{Nakagawa2020,Malnou2021,Valenzuela2020}. However, as discussed below, we describe most of the elements of the model as transmission lines and not as lumped elements. In fact, when comparing different versions, we found that the former described simulations and experimental results more accurately.

Let us consider a unit cell of the CPW artificial line. As depicted in Fig.~\ref{fig:Tmodel}, the unit cell can be divided into several elements, discussed below, to form a circuital model. Then, we can use standard microwave transfer-matrix theory to obtain the transmission matrix of the unit cell which, in turn, allows us to use expressions (\ref{eq:gamma}), (\ref{eq:ZB}) and (\ref{eq:S21}) to determine the properties of the entire line.

Elements I and II are short CPW lines of lengths $\frac{1}{2}l_H$ and $s$, respectively. Their propagation constant and characteristic impedance, needed to form their transmission matrices, can be calculated using theoretical models for superconducting  CPW lines~\cite{Watanabe1994}.

Elements F are CPW stubs of length $l_F$ terminated in a capacitance $C_{end}$. Since they are in parallel, they can be replaced by an admittance equal to two times the admittance
\begin{equation*}
    Y_F = Y_{0F} \frac{Y_{end}+Y_{0F}\tanh{\gamma_F l_F}}{Y_{0F}+Y_{end}\tanh{\gamma_F l_F}},
\end{equation*}%
where $Y_{0F}=1/Z_{0F}$ is the characteristic admittance of the stub, $\gamma_F$ is its propagation constant, and $Y_{end}=j2\pi f C_{end}$. The capacitance $C_{end}$ can be calculated from expressions obtained using conformal techniques~\cite{Yoon2003}.

While studying this model, we realized that, for good accuracy, it was not possible to neglect the effect of the intersection between the central line and the capacitive loads. Although the strong effect of this intersection can be the result of electromagnetic interaction between the load and the central line, we found that two simplified models can be used with good results. One of the models simply considers that the transmission line II of Fig.~\ref{fig:Tmodel} has a length $s+l_L/2$ instead of only $s$. The second option, discussed in the reminder of this article, is to divide the intersection into two free standing superconductor thin sheets. There are three contributions to the impedance per unit length of a free standing superconductor of width $w$~\cite{Tolpygo2022},
\begin{equation*}
    \begin{split}
        \frac{L_c}{l} = & \frac{\mu_0}{8\pi} + \frac{\mu_0}{2\pi} \left( \log{\frac{2 l}{t+w}}+ 0.2235\,\frac{t+w}{l}+\frac{1}{2}  \right) \\
                & +   \frac{\mu_0\lambda^2}{w t}.
    \end{split}      
\end{equation*}%
The first term represents the magnetic field inside the sheet. The second one is its self inductance~\cite{Rosa1908}. The last term is the kinetic inductance per unit length of the superconducting sheet when there is a uniform current distribution through it.

\subsection{Comparison with HFSS Simulations}
Here we present comparisons between the theoretical model described above and the software-aided modelling method~2 presented in Sec.~\ref{sec:hfss2}. First, using the theoretical model we studied the dependence of $Z_0$ and $v/c$ on the the length $w_L$ for the same lines studied in Fig.~\ref{fig:method2b}. For easier comparison, the results are presented in the same figure as solid lines. Both modelling methods show an excellent agreement, especially for values $w_l>\SI{30}{\micro\meter}$.

\begin{figure}[t]%
    \centering
    \subfloat{{\includegraphics[width=0.76\linewidth, clip, trim={0 17.8mm 0 0}]{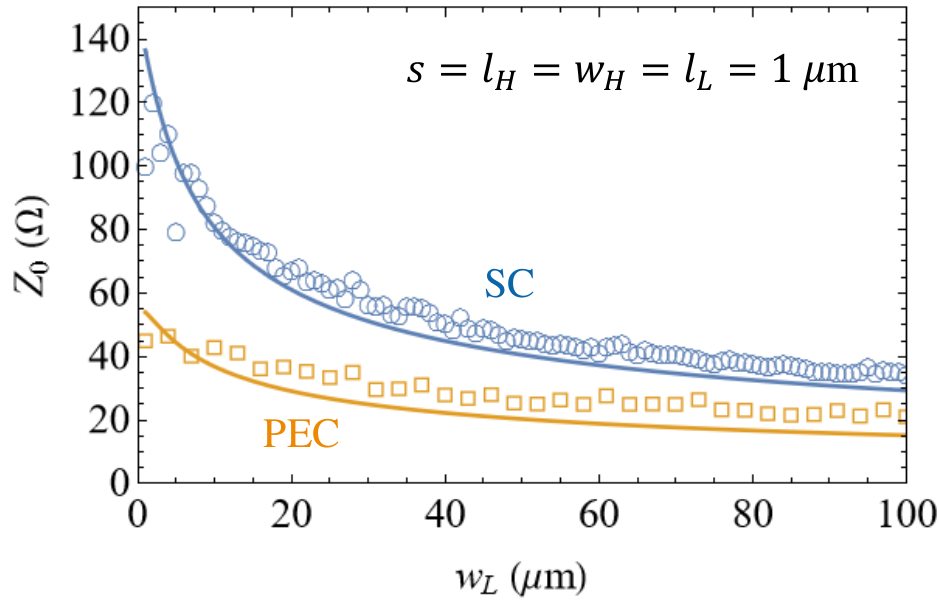} }}\\%
    \vspace{-3.6mm}    
    \subfloat{{\includegraphics[width=.76\linewidth]{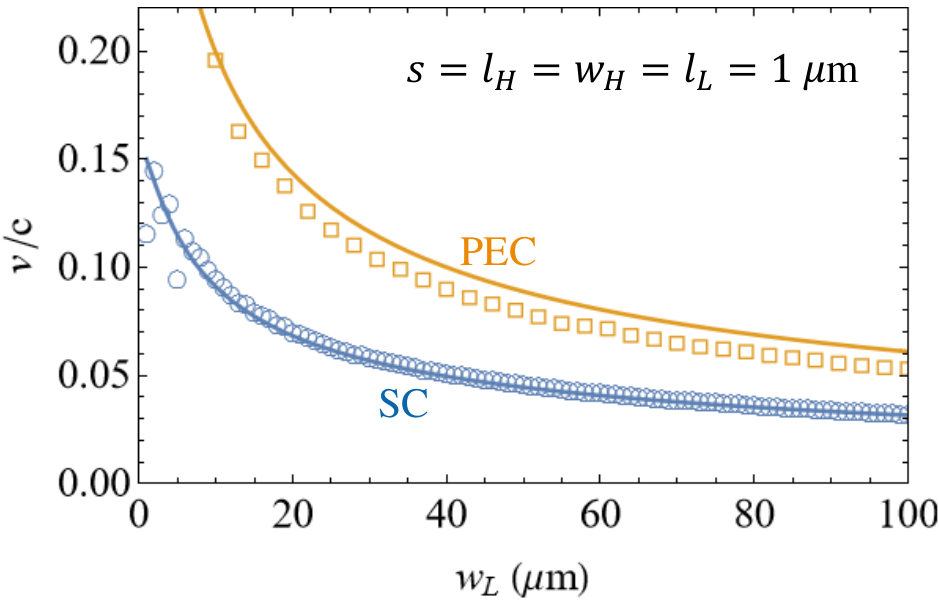} }}%
    \caption{Comparison between software-aided model method~2 (empty symbols) and first-principles model (lines). Both models were applied to lines using a superconductor (SC) and a perfect electrical conductor (PEC). (Top)~Characteristic impedance. (Bottom)~Phase velocity as fraction of the speed of light.}%
    \label{fig:PEC-SC}%
\end{figure}%

For the second comparison, we need first to realize that the first-principles model can be applied without major changes to CPW artificial lines made of a perfect electrical conductor (PEC) instead of a superconductor. Therefore, it can be applied at the same time, without changing any parameter, to both type of lines if they have the same geometrical dimensions. Then, we also applied software-aided modelling method~2 to lines made of PEC. The results and comparison between lines made out of PEC and superconducting material are presented in Fig.~\ref{fig:PEC-SC}. Once more, the agreement is excellent demonstrating the consistency of the theoretical model.

\subsection{Comparison with Measurements}
Using the theoretical model we also calculated the transmission of the Floquet lines that were characterized and presented in Sec.~\ref{sec:Floquet-character}. The results, presented as red solid lines in Fig.~\ref{fig:measure}, show a good agreement with the measurements.

\section{Predicted Performance at High Frequencies}
\begin{figure}[t]%
    \centering
    \subfloat{{\includegraphics[width=0.76\linewidth, clip, trim={0 18.3mm 0 0}]{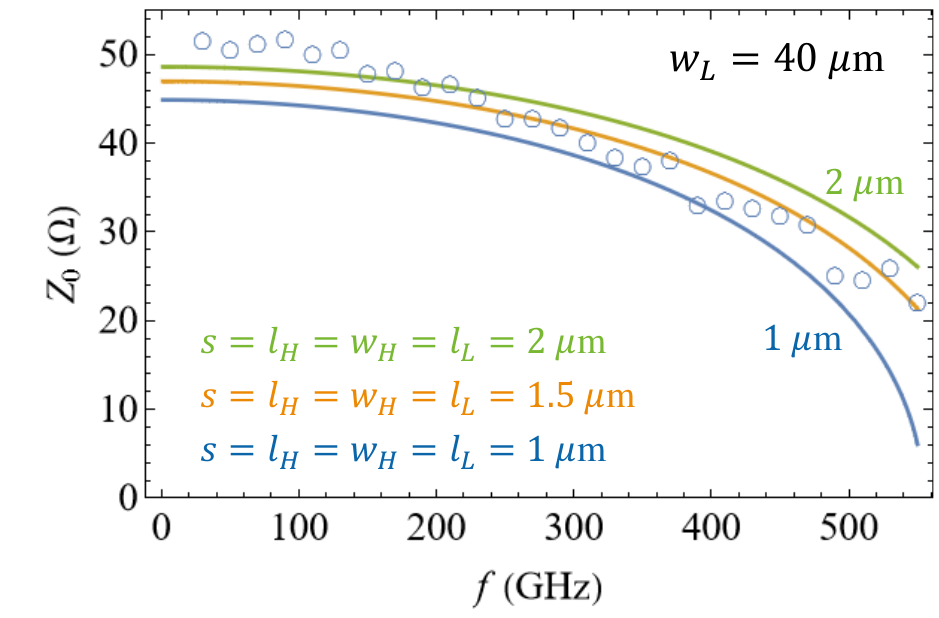} }}\\%
    \vspace{-3.5mm}    
    \subfloat{{\includegraphics[width=.76\linewidth, clip, trim={0 0.9mm 0 0}]{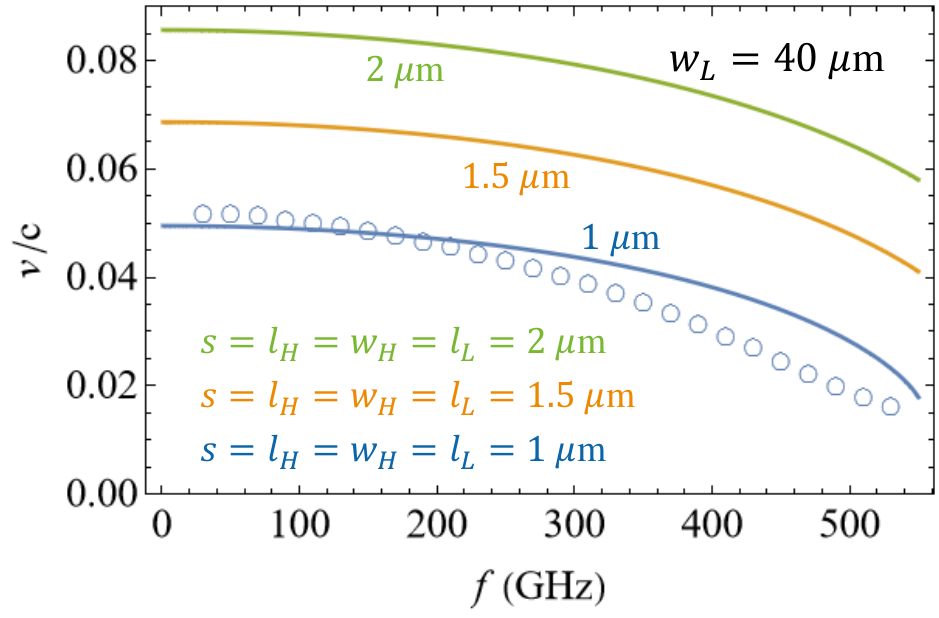} }}%
    \caption{Behaviour of superconducting CPW artificial lines at high frequencies extracted from the first-principles model (lines) and from applying method~2 in windows of \SI{20}{\giga\hertz} (circles). The latter was applied only to an artificial line with $s=\SI{1}{\micro\meter}$. (Top)~Characteristic impedance. (Bottom)~Phase velocity as fraction of the speed of light.}%
    \label{fig:HF}%
\end{figure}%
Assuming a single mode propagation, the theoretical model presented in previous section can predict the performance of superconducting CPW artificial lines at high frequencies. The results are presented in Fig.~\ref{fig:HF} (solid lines). We can see that if no other mode is excited, the propagating mode shows small dispersion up to a few hundred~\si{\giga\hertz}. These results are corroborated by applying method~2 (\ref{sec:hfss2}), in windows of \SI{20}{GHz}, in order to obtain the wavenumber and characteristic impedance of an the artificial line with  $s=\SI{1}{\micro\meter}$ (Fig.~\ref{fig:model}b). The calculated values, presented as circles in Fig.~\ref{fig:HF}, agree fairly well with the theoretical model. 

\begin{figure}[!t]
    \centering
    \includegraphics[width=0.8\linewidth]{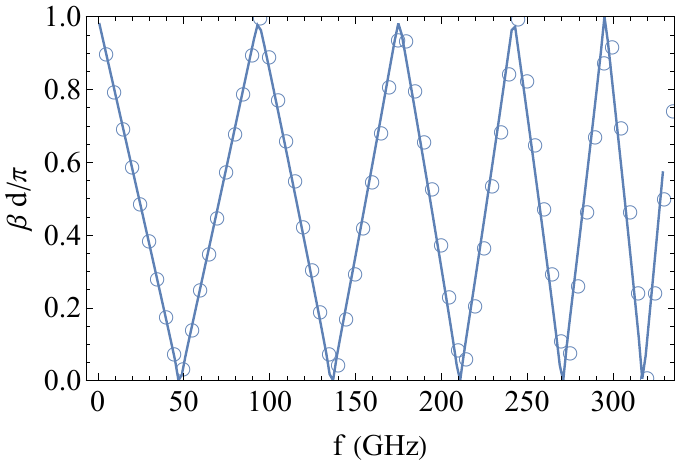}
    \caption{Wavenumber of the an artificial line with $s=l_H=l_L=w_H=\SI{1}{\micro\meter}$ and $w_L=\SI{40}{\micro\meter}$. The calculations were made using the MTM method (line) and method~2 described in this work (circles). The latter data correspond to those presented in Fig.~\ref{fig:HF}.}
    \label{fig:MTMM}
\end{figure}%
For the purpose of determining the influence of other modes, we applied the Multimodal Transfer Matrix (MTM) method~\cite{Mesa2021,Giusti2022} to the same artificial line. Two modes, identified as symmetric and asymmetric CPW, were used in the calculations. The resulting wavenumber is presented as the solid line of Fig.~\ref{fig:MTMM}. Up to the highest simulated frequency, the influence of the antisymmetrical mode was found to be minimal. Moreover, the results coincide very well with the simulation method~2 (open circles), confirming that the transmission is monomodal with small dispersion up to few hundred~\si{\giga\hertz}. These features make the artificial CPW line a good candidate to implement superconducting parametric amplifiers at high frequencies.


\section{Conclusions}
We have presented two strategies for designing superconducting  artificial lines and Floquet lines made out of them. The first strategy combines the use of electromagnetic simulation software with standard transfer-matrix theory. The former is used to extract the properties of the artificial line that, in turn, allow the use of the latter to describe Floquet lines. In contrast with other methods, it requires much less computational resources since it does not require the simulation of the entire unit cell of a Floquet line. The second strategy uses first principles to construct an equivalent circuit model. Furthermore, these strategies were compared against each other showing excellent agreement. We have also demonstrated that both of them describe very well the measured transmission of superconducting Floquet lines fabricated using NbTiN. One of the main advantages of using the design strategies presented in this work is that they allow a fast modelling of Floquet lines made out of CPW artificial lines, which is important in, for example, the design of travelling-wave parametric amplifiers. Finally, using the first-principles model we showed that the CPW artificial line is a good candidate to implement parametric amplifiers at frequencies of few hundred~\si{\giga\hertz}.

\section*{Acknowledgments}
F.~P.~Mena acknowledges partial support from ANID through Fondecyt grant 1180700 and thanks C.~Jarufe for fruitful discussions. R.~Finger acknowledges the support of ANID through its funds Basal FB210003, Fondecyt 1221662 an Fondef ID21-10359. The work of D. Valenzuela was supported by grant DOCTORADO BECAS CHILE 2020 – 21200705. F. Pizarro acknowledges partial support from ANID through Fondecyt grant 1221090.  The authors thank SRON and the Kavli Institute of Nanoscience Delft for the use of their clean room facilities for device fabrication.

\bibliographystyle{IEEEtran}
\bibliography{references.bib}



\vfill

\end{document}